\documentclass{iserc}
\usepackage{soul}
\usepackage{multirow}
\usepackage[colorinlistoftodos]{todonotes}

\conference{Proceedings of the 2019 IISE Annual Conference\\
H.E. Romeijn, A. Schaefer, R. Thomas, eds.}

\title{\titlesize Multivariate Modeling for Sustainable and Resilient Infrastructure Systems and Communities}
\author{
Renee Obringer\\Environmental and Ecological Engineering\\Purdue University\\West Lafayette, IN 47901, USA\\
\vspace{0.3cm}
Roshanak Nateghi\\Industrial Engineering\\Purdue University\\West Lafayette, IN 47901, USA}
\authorlist{Obringer and Nateghi}
\abstractID{610237}

\begin{document}
\begin{large}

%
%
%
\end{large}
\newpage
\maketitle

\begin{abstract}
Sustainability and resilience of urban systems are multifaceted concepts, requiring information about multiple system attributes to adequately evaluate and characterize. However, despite the scientific consensus on the multivariate nature of these concepts, many of the existing techniques to model urban sustainability and resilience are unidimensional in nature, focusing on a characterizing a single element of highly interconnected urban systems. We champion a paradigm shift in modeling urban sustainability and resilience, using an integrated approach to simultaneously estimate multiple interconnected (correlated) system attributes of sustainability and resilience as a function of key environmental factors. We present a novel case study and review a few recent studies to illustrate the applicability and benefits of the multivariate approach to modeling urban sustainability and resilience. Our proposed framework can be utilized by infrastructure managers, urban planners, and researchers to conceptualize and assess urban sustainability and resilience more holistically, and to better understand the key factors in advancing the sustainability and resilience of infrastructure systems.
\end{abstract}

\section*{Keywords}
Statistical Learning, Urban Sustainability and Resilience, Multivariate Modeling, Integrated Modeling

\section{Introduction}
Urban areas currently contain over 50\% of the world's population, and are expected to grow to hold nearly 70\% of the population by 2050 \cite{TheWorldBank2010}. This unprecedented growth will put stress on urban infrastructure systems, which will only be exacerbated by climate change. In fact, many cities around the world have already begun to feel the stress brought on by aging infrastructure, growing urban populations, more frequent weather and climate extremes, and increasingly difficult budgetary constraints. In fact, it is expected that updates to our current infrastructure systems (including the telecommunication, road, rail, water, and electricity systems) will cost upwards of \$70 trillion (USD) by 2030 \cite{OrganizationforEconomicCo-OperationandDevelopment2007}. If one considers that urban areas will continue to grow in size and population, while facing unprecedented climate variability and change, it becomes increasingly apparent that cities must work to improve their sustainability and resilience, so that they may be prepared for the challenges the next few decades will bring.\\

Despite significant research progress in modeling infrastructure resilience in recent decades, fundamental research gaps remain. These gaps are rooted in the failure to conceptualize the multidimensional resilience of systems. Specifically, although it is well documented that urban resilience itself is multifaceted and complex, requiring a systems approach to analyze \cite{Meerow2016a}, much of the research continues to focus on a single aspect of the overall concept of urban resilience. It is likely that this fragmented approach has led to incomplete information about the state of the system, leading to sub-optimal resilience investment strategies and escalating disaster losses. In recent years, there has been a push for more interdisciplinary research within the field, with several researchers calling for new paradigms that allow for more complex studies of urban resilience \cite{Pandit2015,Ferrer2018}. However, these calls for paradigm shifts are often very abstract and conceptual in nature and do not provide concrete means of assessing the multivariate (and correlated) attributes of systems, and thus fall short of guiding researchers in ways in which they can practice these new paradigms for studying urban sustainability and resilience. Here, we present a novel framework, leveraging recent developments in statistical machine learning, to evaluate the multiple aspects of resilience simultaneously, and thus better characterize urban sustainability and resilience as a whole. Our proposed framework hinges on first identifying multiple performance metrics (i.e., response variables) that are hypothesized to best characterize the state of sustainability/resilience of the system of interest, based on inputs from experts and stakeholders, and then use the multivariate tree boosting algorithm \cite{Miller2016} to estimate the joint distribution of the multiple performance metrics. \\

The boosting meta-algorithm is a sequential learning process that can improve the the performance of a given classifier by re-weighting the training samples and giving higher weights to poorly predicted data points, such that they can be better predicted in the next iteration \cite{Friedman2001}. The final estimate is then based on aggregation of the predictions across all the classifiers, with higher weights given to the better classifiers. Multivariate tree boosting applies the boosting meta-algorithm to multivariate regression trees; the covariance structure of the multiple response variables can then be harnessed to make simultaneous predictions of the multivariate response as a function of a wide suite of independent variables \cite{Miller2016}. To identify the key predictors of the multivariate performance measures of systems, the relative influence of a given independent variable can be characterized by measuring the reduction in prediction error due to any split on that predictor, summed over all trees in the model. The estimated reductions in prediction errors attributed to each independent variable can then be leveraged to rank the predictors according to their relative influence. In the case of multidimensional response variables, the univariate relative influence is first calculated for each predictor variable as well as each of the response variables. Summing the importance over all response variables establishes a global importance measure for the predictor across all the response variables. In addition, to facilitate inferencing based on the multivariate tree boosting, the relative influence measures can be grouped by first calculating the distance (e.g., Manhattan distance) between the predictors and the pairs of response variables, respectively. Independent variables that explain similar covariance trends in the response variables will be closer together as will pairs of dependent variables that are functions of a similar subset of predictor variables. The estimated distance matrices can be used to group the predictors that explain covariance in similar pairs of response variables, and the pairs of responses that are dependent on similar subsets of independent variables via hierarchical clustering. Below, we apply the proposed framework to three distinct case studies. The studies discussed below demonstrate the benefits of multidimensional modeling of urban infrastructure resilience---i.e., the power of harnessing the covariance structure of interdependent performance metrics to characterize systems sustainability and resilience more holistically. \\

Below we discuss a novel study that utilizes the above-mentioned framework to evaluate urban infrastructure resilience, as well as review a few recent applications as the framework.

\section{Modeling Community Resilience Facing Tsunami Impacts in Japan}
Tsunamis are incredibly costly disasters, both in terms of human lives and infrastructure damage. Typically, the damage caused by these events is measured as a function of the number of physical structures damaged or the mortality rates. These variables are often considered as isolated measures in predictive modeling. One such study by Nateghi et al. (2016) was performed to assess the accuracy of various predictive algorithms a to estimate tsunami damage and death rates, separately \cite{Nateghi2016}. The authors demonstrated that the ensemble-of-trees predictive algorithms were able to reasonably estimate both measures, but damage rates were better predicted than death rates \cite{Nateghi2016}. We hypothesized that the discrepancy in the accuracy of the predictions of the two measures could be remedied by analyzing the data using a multivariate algorithm that allows for multiple, interdependent response variables. We implemented our framework based on multivariate tree boosting to test this hypothesis. \\

The data for this case study was originally obtained from a variety of sources as outlined in Nateghi et al. (2016) \cite{Nateghi2016} and was made publicly available. The data includes various information for select municipalities in the Miyagi and Iwate prefectures in Japan that were heavily impacted by the 1896, 1933, 1960, and 2011 tsunamis. The response variables were damage rate and death rate. The predictor variables included the year of the tsunami, the population before the tsunami, the number of dwellings before the tsunami, the area of the municipality, the coastal forest area, the maximum tsunami runup, the presence of bay mouth breakwater, the sea wall height (minimum and maximum), the flooded area, the topography, and the prefecture. \\

The results of applying the multivariate methodology is depicted in the figure below. Figure \ref{fig2}
demonstrates the benefits of using the multivariate framework, as the prediction of the death rate greatly improved (compared to Nateghi et al. (2016)), while the prediction of the damage rate remained the same. The out-of-sample error (mean squared error) for the damage and death rate predictions were 0.31 and 0.27, respectively. \\ 
\begin{figure}[!h]
\centering
\includegraphics[width=3.5in]{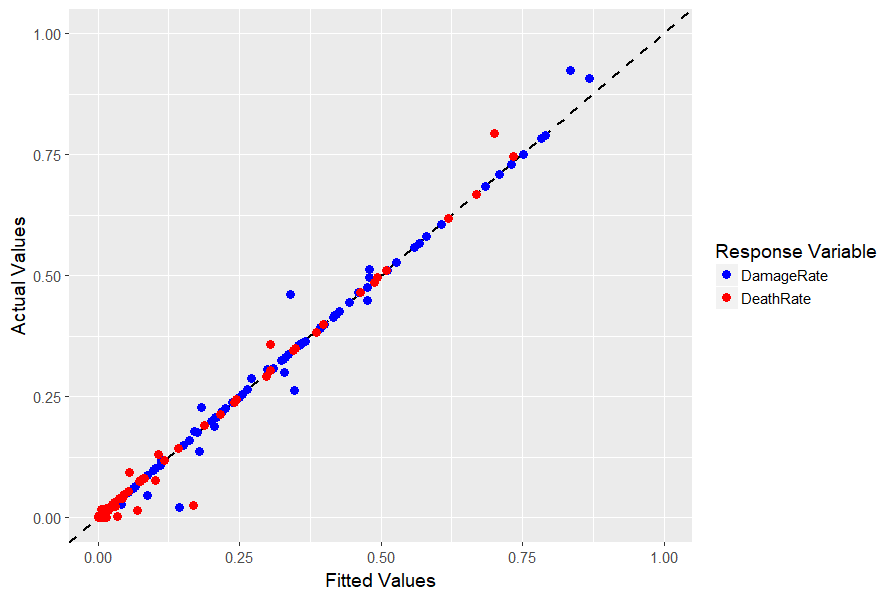}
\caption{Plot of the observed values versus the estimated values for the damage rate (blue) and the death rate (red). Note that the death and damage rates have been normalized to be depicted on the same graph }
\label{fig2}
\end{figure}

\begin{figure}[!h]
\centering
\includegraphics[width=4.5in]{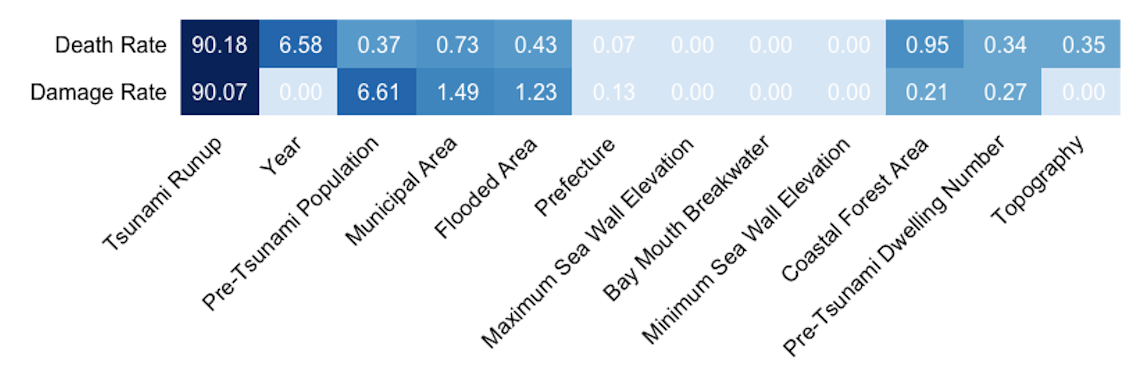}
\caption{Relative importance of the predictor variables for predicting each interdependent response variable.}
\label{fig3}
\end{figure}

As discussed earlier, one of the benefits to our framework is the ability to determine the important variables in the model for multiple response variables simultaneously. The results of this study show that the most important variable for predicting both damage and death rates is the maximum tsunami runup. The runup is often considered one of the most damaging forces of a tsunami, therefore it is logical that it should be the most important predictor for damage and death rates (see Figure \ref{fig3}). However, in the univariate study by Nateghi et al., the year was found to be the most important predictor of death rates, with tsunami runup falling to second \cite{Nateghi2016}. \\ 

Moreover, the analysis suggests that when considering death and damage rates concurrently, seawall heights might not be among the key predictors of impact as previously suggested.
In short, the mulitvariate framework improves not only the accuracy of the prediction, but offers different insights than the univariate model. 

\section{Discussion}
The results of this study demonstrate the viability of our framework, which allows users to leverage interdependencies in the response data to improve predictions and better understand the key factors affecting urban sustainability and resilience. Specifically, our framework improved upon previous work that focused on predicting tsunami damage and death rates as univariate response variables. In the previous work, death rates were poorly predicted \cite{Nateghi2014}, but when we implemented our integrated framework, we were able to get improved predictions for death rates while maintaining the accurate predictions of damage rates found in the previous study. In addition to this study on tsunami damage, we have recently applied our framework to other aspects of urban sustainability and resilience, namely the water-energy nexus \cite{Obringer} and electrical power grids \cite{Nateghi2018}. Those studies confirm our results, and further demonstrate the power of this framework to greatly improve our models. 

\subsection{Modeling the Water-Energy Nexus}
One of the many interdependencies present in urban systems is between water and energy, known as the water-energy nexus \cite{Hussey2012}. The nexus has been documented in both the supply side, that is the water needed during electricity generation and the electricity needed during water treatment and distribution \cite{Sovacool2009,Sanders2012}, as well as the demand side, meaning the interdependent use of water and electricity within the residential sector \cite{Ruddell2014}. However, when evaluating the water or electricity needed by a city, researchers rarely take these interdependencies into account. Not accounting for the interdependencies is particularly precarious when predicting the future water or electricity demands, as the actual amount of water available or needed will depend on the amount of electricity needed. Moreover, by accounting for climate change in making future projections of the water and/or electricity sectors, it becomes increasingly apparent that accurate models that consider the interdependencies between the systems are necessary. Obringer and Nateghi (2018) used the multivariate framework to model the climate-sensitive portion of residential water and electricity use in six Midwestern cities \cite{Obringer}.\\

Data was collected for the cities of Chicago (IL), Cleveland (OH), Columbus (OH), Indianapolis (IN), Madison (WI), and Minneapolis (MN). The response variables included monthly water use and monthly electricity use, both normalized by the number of customers served. The independent predictor variables included: average maximum dry bulb temperature, average dew point temperature, average relative humidity, average wind speed, and the El Ni\~no/Southern Oscillation (ENSO) strength. It was the focus of this study to predict the portion of the water-energy nexus that was sensitive to climate, thus only climatic variables were used in the model.\\ 

\begin{figure}[!h]
\centering
\includegraphics[width=3.5in]{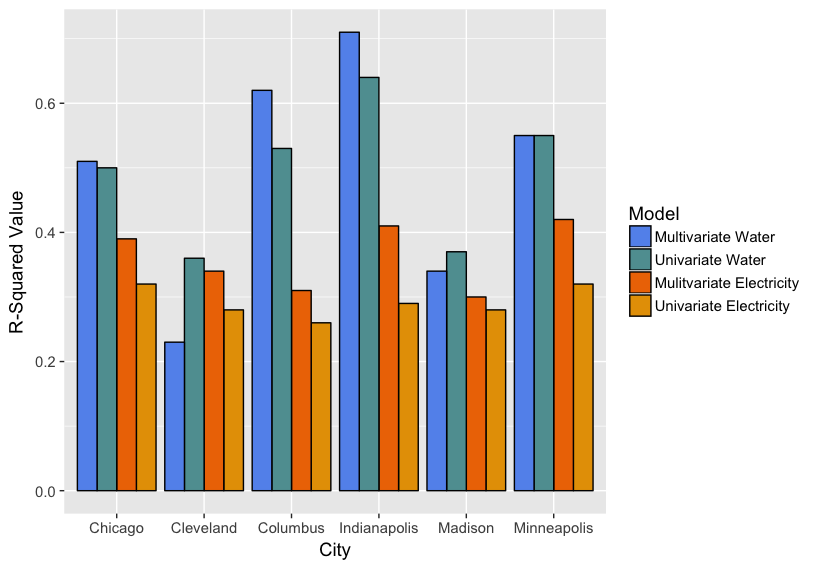}
\caption{Plot of the out-of-sample $R^2$ values for each response variable in the multivariate and univariate models. \cite{Obringer}}
\label{fig1}
\end{figure}

Following the analysis, it was found that the multivariate model accurately predicted the climate-sensitive portion of the water and electricity demand in the six cities, based on both out-of-sample root mean squared error and goodness of fit measures. In fact, in the majority of the cities, the multivariate model was able to explain more variance in the response data than the univariate model, as shown in Figure \ref{fig1}, which depicts the out-of-sample $R^2$ for both response variables in the multivariate and univariate models \cite{Obringer}. Since the univariate model was the same basic algorithm as the multivariate model (i.e., gradient tree boosting), the only difference between the model runs was the consideration of interdependencies in the multivariate model, thus indicating the importance of such interdependencies. These results show that it is important to consider the interconnectivity between water and electricity use, even if one is only interested in a single response variables, as it improves the overall predictive accuracy.

\subsection{Modeling the Resilience of an Electrical Power Grid}
As mentioned in the introduction, resilience is an inherently multidimensional concept. For instance, when an electrical power grid is affected by a hurricane, the system resilience is often measured by three different, yet connected aspects: the number of damage assets or number of outages \cite{Nateghi2014,Nateghi2014a}, system restoration lengths or outage duration \cite{Liu2007,Nateghi2011}, and number of customers affected \cite{Staid2014, Guikema2014}. However, these variables are often considered stand alone measures of resilience in the research, which may lead to a misrepresentation of the system's overall resilience. The case study below illustrates how the multivariate tree boosting algorithm can be leveraged to evaluate the resilience of an electrical power grid affected by Hurricane Katrina using all three measures simultaneously \cite{Nateghi2018}.\\

The data collected in this study consisted of three response variables and 63 predictor variables. The response variables were the measures of resilience discussed previously: outage counts, outage duration, and number of customers affected. The predictor variables described the characteristics of the hurricane, the system typology, and the characteristics of the service area, including topography, climate, and land use \cite{Nateghi2018}.\\

The results of this study showed that the multivariate model accurately predicted the various measures of resilience, and was significantly better than the null, or mean-only, model \cite{Nateghi2018}. One of the benefits of the multivariate model is the ability to analyze the impacts of various scenarios on the response variables---a tool that is especially helpful in resilience studies. In this case study, we compared the ability of two methods to reduce tree damage to power lines: tree-trimming and under-grounding. The results show that the benefits of each method are different for the the different resilience measures. For example, if one considers the number of customers affected to be the measure of resilience, under-grounding would lead to 79\% less inoperability of the system, while tree-trimming would only lead to 11\% less \cite{Nateghi2018}. Conversely, if one considers outage duration to be the resilience measure, under-grounding results in an 89\% reduction and tree-trimming results in a 68\% reduction \cite{Nateghi2018}. This demonstrates the importance of including multiple measures of resilience in models, as only including one could lead to a biased understanding of what strategies would be beneficial for a given system. 

\section{Conclusion}
Around the world, urban areas are growing at unprecedented rates and can be expected to continue this growth throughout the next few decades \cite{TheWorldBank2010}. In light of this expected growth, in combination with future climate change, it is imperative that cities work towards improving their resilience quickly. In order to facilitate this improvement, researchers must provide accurate predictive models that can aid in cities preparation. However, as most models used today are univariate, it is possible that accuracy is being lost. That being said, recent developments in statistical learning theory can be leveraged to improve model accuracy by including multiple response variables in predictions. In this study, we detailed three studies that utilize the algorithm known as multivariate tree boosting to predictive various aspects of urban resilience. The first was a new analysis that built off a previous study on predicting damage and death rates of tsunamis in Japan. The results showed an improvement over the univariate model when implementing our mulitvariate framework. The following two studies were reviews of recent studies that implemented the framework. The first of these focused on two interdependent systems within urban areas---water and electricity use. By considering these systems as two interconnected response variables, the predictive accuracy was improved \cite{Obringer}. The next study demonstrated the ability of the algorithm to be used when predicting multiple resilience measures within a single urban system. The study focused on predicting three measures of electrical grid resilience following a hurricane, and demonstrated that different measures could lead to different conclusions regarding the viability of various preparatory actions \cite{Nateghi2018}. Overall these studies indicate that there is knowledge to be gained from adopting a more integrative approach to resilience modeling---knowledge which would have been lost had a traditional, univariate approach been taken.

\section*{Acknowledgements}
The authors would like to acknowledge the Purdue Climate Change Research Center (PCCRC), the Center for Environment (C4E) for their support as well as the NSF grants \#1826161 and \#1728209.

\bibliographystyle{ieeetr}
\small{\bibliography{iisepaper}}

\end{document}